\documentclass[12pt,a4paper]{article}
\usepackage[a4paper,width=150mm,top=25mm,bottom=25mm]{geometry}
\usepackage{graphicx} % Required for inserting images
\usepackage{color}
\usepackage{physics}
\usepackage{amsmath}
\usepackage{amssymb}
\usepackage{amsthm}
\usepackage{authblk}
\usepackage[normalem]{ulem}
\usepackage{tikz}
\usepackage{adjustbox}
\usepackage{url}
\usepackage{quantikz}
\usepackage[utf8]{inputenc}
\title{Quantum MeanFlow: single-shot generative sampling on NISQ hardware}
\author[1,2]{Ashish Joshi}
\author[1,3]{Eshaan Mistry}
\author[1,2]{Takahiko Koyama\thanks{Corresponding author: \texttt{takahiko.koyama@keio.jp}}}
\affil[1]{Human Biology-Microbiome-Quantum Research Center (WPI-Bio2Q), Keio University, Tokyo, Japan}
\affil[2]{Keio University Sustainable Quantum Artificial Intelligence Center (KSQAIC), Keio University, Tokyo, Japan}
\affil[3]{University of California, Berkeley, CA, USA}
\date{\vspace{-2.5em}}

\begin{document}
\maketitle
\begin{abstract}
Quantum generative models offer a promising framework for exploring whether quantum computation can enhance generative machine learning.
Flow matching is a generative method in which samples are generated by transporting a simple, known distribution to the target data distribution with a learned velocity field. 
Its quantum counterpart, known as quantum flow matching (QFM), was introduced recently, and, like its classical counterpart, requires integrating an ordinary differential equation over many time steps during inference. 
As each step requires the output from the previous step, the circuit submission is sequential and a drawback on quantum computers as they have high input/output costs. 
To alleviate this problem, we introduce Quantum MeanFlow (QMF), the quantum analogue of the MeanFlow formulation, which allows single-step sample generation. 
While the QFM learns an instantaneous velocity field at each time step, QMF learns the average velocity over a time interval. 
We use a parameterized quantum circuit to learn these velocity fields and benchmark the two methods on the MNIST dataset.
We show that while single-step QMF has lower image quality compared to multi-step QFM, it performs better than the single-step QFM sampling at every shot count.
Both of our models are executed on IBM quantum computers and best-of-$N$ rejection sampling recovers most of the accuracy lost to device noise without modifying the circuit.
This is especially advantageous for QMF which has only one circuit evaluation per image.
Here, We establish QMF as a viable method for single-step quantum generative sampling, saving on quantum circuit evaluations per generated sample.
\end{abstract}

\section{Introduction}
\label{s_introduction}

Generative machine learning attempts to learn the underlying probability distribution from a dataset and generate samples from the learned distribution. 
The modern landscape of gerenerative machine learning started with variational autoencoders (VAEs) \cite{kingma2014vae}, generative adversarial networks (GANs) \cite{goodfellow2014gan} and auto-regressive models \cite{oord2016pixelrnn}. 
Diffusion models \cite{sohldickstein2015diffusion,ho2020ddpm,song2021sde} improved on several aspects of the previous methods, providing training stability and the ability to generate high-quality data \cite{dhariwal2021beatgans}.
Diffusion models learn to denoise the data step-by-step and can be described by stochastic differential equations \cite{song2021sde}. 
An equivalent but mathematically simpler formalism is flow matching, which is instead described by ordinary differential equations \cite{lipman2023flow,liu2023rectified,albergo2023interpolants,holderrieth2025intro}. 
The model learns the probability flow between the noise and data distribution, and once trained, inference is much faster because a simple ordinary differential equation (ODE) is integrated to generate samples.\\

Alongside the developments described above, the disparate field of quantum computing is getting much attention. 
Quantum computing uses the principles of quantum mechanics as the basis of computation \cite{nielsen2010quantum} and has the potential to provide speedups over classical algorithms for specific classes of problems \cite{shor1997factoring,grover1996search}. 
Since machine learning and deep learning methods often have high computational costs, it is natural to consider if quantum computing can provide any benefits for these methods.
The relatively new subfield of quantum machine learning (QML) \cite{biamonte2017qml,cerezo2021vqa} aims to explore the potential of quantum computing in machine learning, with many interesting results over the past few years. 
Even without an established quantum advantage, implementing generative models on quantum hardware provides a framework for investigating whether quantum circuits can offer useful alternative representations of complex data distributions.
Quantum models for generative machine learning have been gaining interest recently, with models for quantum GAN \cite{lloyd2018qgan,dallairedemers2018qgan,zoufal2019qgan,huang2021qganimage}, quantum VAE \cite{khoshaman2019qvae,romero2017qautoencoder}, quantum diffusion \cite{zhang2024quddpm,kolle2024qddm,cacioppo2023qdm,defalco2024qldm,kwun2025msquddpm} and quantum flow matching \cite{cui2025qfm,hahm2024qfm}. 
While quantum diffusion models are being explored for both quantum and classical data, quantum flow matching (QFM) has only been explored in the context of quantum data. 
The closest work on classical data is a hybrid quantum-classical normalizing flow \cite{zhu2024hqcnf}, which learns an invertible map rather than a velocity field.
Here, we fill this gap by focusing on QFM for generating classical data. \\ 

As we are still far from the era of fault-tolerant quantum computers, most of the QML models are implemented either in simulations or on the currently available noisy intermediate scale quantum (NISQ) devices \cite{preskill2018nisq}. 
However, training a quantum circuit on NISQ devices is not feasible because of high rate of errors, in addition to the vanishing gradients that affect deep variational circuits \cite{mcclean2018barren}. 
Instead, we perform training classically and use the trained quantum models to generate samples on a real device. 
This still requires sampling for several time steps on a quantum computer, with the output of one step being the input for the next. 
As data input and output on a quantum computer is expensive \cite{benedetti2019pqc,cerezo2021vqa}, the cost of doing this increases with number of steps, which is often required to generate good quality images. 
Recent work has tackled this problem by proposing single shot sampling in classical flow matching. 
This method is called MeanFlow \cite{geng2025MeanFlow} and it was shown that it outperforms all the other single and few shot sampling methods \cite{song2023consistency,frans2025shortcut,salimans2022progressive}.\\

In this work, we present the first QFM model with MeanFlow for classical image generation. We call this Quantum MeanFlow (QMF). 
We train both QFM and QMF models on the MNIST dataset in simulation and perform inference on quantum hardware. 
We show that the QMF model can perform single-step sampling on the hardware, requiring only one circuit evaluation to generate one image. 
We also observed that at lower number of shots, multi-step sampling in both QFM and QMF models can alleviate the effects of sampling noise.
Our results establish single-step QMF as a viable approach for quantum generative sampling that eliminates the sequential circuit evaluations required by multi-step QFM, making it better suited to the constraints of current NISQ hardware.\\

The paper is organized as follows. 
In section~\ref{s_background}, we give the background of classical flow matching and MeanFlow formulations.
In section~\ref{s_qmf}, we present our QMF formulation, with data processing steps and classical methods in section~\ref{ss_data}, the quantum ansatz for learning the velocity fields in section~\ref{ss_ansatz} and training and sampling procedure in section~\ref{ss_training}.
The results are presented in section~\ref{s_results} with discussion in section~\ref{s_discussion}.
%\textcolor{red}{one job - two jobs. }
%\textcolor{red}{overall flow and story of the manuscript}
%CHECK ALL REFERENCES IF VALID

\section{Background}
\label{s_background}
\subsection{Classical flow matching}
\label{ss_flowmatching}

Flow matching is a framework for generative modelling in which samples are generated by transporting a simple, known distribution onto the unknown data distribution \cite{lipman2023flow,liu2023rectified}. 
The object being learned is a time-dependent velocity field $v_\theta(\cdot,t)$, where $\theta$ are the parameters to be learned. 
This velocity field defines a flow between the reference distribution and the data distribution through the ordinary differential equation
\begin{equation}
\tfrac{\mathrm{d}}{\mathrm{d}t}\psi_t(z)=v_\theta(\psi_t(z),t),
\end{equation}
where $z$ are the samples and $\psi_0(z)=z$. The solution to the above ODE is defined by the trajectory $\psi_t(z)$, which gives the sample at time $t$.
Throughout this paper we follow the convention of \cite{geng2025MeanFlow}, in which
$t=0$ denotes data and $t=1$ noise\footnote{This reverses the convention of
\cite{lipman2023flow,liu2023rectified}, where time runs from noise to data. We adopt the
former so that the definitions of Sec.~\ref{ss_MeanFlow} can be used directly.}.
Usually, the reference distribution is a Gaussian $\mathcal{N}(0,I)$. Let $p_t(z)$ be the probability density of the samples at time $t$, so that $p_0$ is the data distribution and $p_1=\mathcal{N}(0,I)$ is the reference distribution. These marginal densities interpolate between the two endpoints and evolve according to the continuity equation \cite{holderrieth2025intro},
\begin{equation}
\frac{\partial p_t(z)}{\partial t} + \nabla_z\!\cdot\!\big(p_t(z)\,v(z,t)\big)=0,
\label{eq:continuity}
\end{equation}
where $\nabla_z$ denotes the divergence with respect to $z$. 
This equation expresses conservation of probability mass transported by the velocity field $v(z,t)$.
During inference, the ODE is solved in the reverse direction from $t=1$ to $t=0$ to generate samples. 
However, the velocity describing the required interpolation between the two distributions is not known.
In flow matching, this is overcome by fixing the interpolation and then learning the velocity field that might generate it.
The velocity field can then be obtained by regression on individual pairs of a data sample and a noise sample for which the target velocity is known exactly. 
Thus, the velocity field is learned without ever simulating the ODE.
We follow the method as described in \cite{holderrieth2025intro} and refer the readers to the same for more details.\\

We use the linear interpolation in which a data sample $x$ and a noise sample $\varepsilon\sim\mathcal{N}(0,I)$ are joined by the straight segment
\begin{equation}
z_t=(1-t)x+t\varepsilon ,
\end{equation}
with a constant velocity $\varepsilon-x$.
Introducing conditioning to the class label $c$ in the velocity field and sampling $t\sim\mathcal{U}(0,1)$, the flow matching loss $\mathcal{L}_{\mathrm{FM}}(\theta)$ can be written as
\begin{equation}
\mathcal{L}_{\mathrm{FM}}(\theta)=
\mathbb{E}_{t,x,\varepsilon}\big\|\,v_\theta(z_t,t,c)-(\varepsilon-x)\,\big\|^2.
\label{eq:fm}
\end{equation}
Since the data samples $x$ and the noise samples $\varepsilon$ are drawn independently, many of these pairs may pass through the same point $z_t$ with different velocities. 
The trained velocity field may only represent their average, which makes curved trajectories during generation. Thus, the ODE must be integrated in several small time steps.
This makes generation sequential, as the output of the current step is used as the input for the next step, and the cost of generating samples grows with number of steps. 
To reduce this sampling cost, recent works have proposed some modifications of the vanilla flow matching in rectified flow, shortcut models and few-shot models \cite{liu2023rectified,song2023consistency,frans2025shortcut,geng2025MeanFlow}. 
In the next section, we describe the MeanFlow model, that attempts to generate data in a single step.

\subsection{Classical MeanFlow}
\label{ss_MeanFlow}

Flow matching, as described above, learns instantaneous velocity at each time step. 
MeanFlow (MF) replaces this by learning an average velocity over a time interval. This makes single-step sampling possible.
We use the same notation as in the previous section: $z_t=(1-t)x+t\varepsilon$ with $t=0$ at the data sample $x$ and $t=1$ at the noise sample $\varepsilon$, with the instantaneous velocity along that segment as $v=\varepsilon-x$.
For two time points $r$ and $t$, with $r\leq t$, the average velocity $u$ over the interval $[r,t]$ is given by
\begin{equation}
u(z_t,r,t)=\frac{1}{t-r}\int_{r}^{t} v(z_\tau,\tau)\,\mathrm{d}\tau.
\end{equation}
The relationship between the average velocity, initial time, final time and displacement $z_t-z_r$ is given by
\begin{equation}
(t-r)\,u(z_t,r,t)=z_t-z_r.
\label{eq:avgvel}
\end{equation}
This equation is exact for all interval lengths. Therefore, if $u$ is learned correctly, the model can move from one end of the trajectory to the other in a single evaluation. 
In other words, during generation, $z_1\sim\mathcal{N}(0,I)$ is drawn and $z_0=z_1-u_\theta(z_1,0,1,c)$ is produced, $u_\theta$ is the learned average velocity field and $c$ is the class label.
Thus, the many small steps of Sec.~\ref{ss_flowmatching} are replaced by one large jump.\\

Since $z_r$ is not known in advance during generation, Eq.~\eqref{eq:avgvel} cannot be used directly for training. Differentiating Eq.~\eqref{eq:avgvel} with respect to $t$
\begin{equation}
u(z_t,r,t)=v(z_t,t)-(t-r)\,\frac{\mathrm{d}}{\mathrm{d}t}u(z_t,r,t),
\qquad
\frac{\mathrm{d}u}{\mathrm{d}t}=v\cdot\partial_z u+\partial_t u ,
\label{eq:identity}
\end{equation}
where the second expression is the total derivative of $u$ along the trajectory. This equation is known as the MeanFlow identity and can be used to write down the MF loss $\mathcal{L}_{\mathrm{MF}}(\theta)$
\begin{equation}
\mathcal{L}_{\mathrm{MF}}(\theta)=
\mathbb{E}_{x,\varepsilon,r,t}\big\|\,u_\theta(z_t,r,t,c)-\mathrm{sg}\!\left[\,v-(t-r)\,
\tfrac{\mathrm{d}}{\mathrm{d}t}u_\theta\,\right]\big\|^2 ,
\label{eq:mfloss}
\end{equation}
where $\mathrm{sg}[\cdot]$ denotes a stop-gradient, meaning the expression inside the brackets is not differentiated when updating $\theta$.

\section{Quantum MeanFlow (QMF)}
\label{s_qmf}

\begin{figure}
        \centering
        \includegraphics[width=1\linewidth]{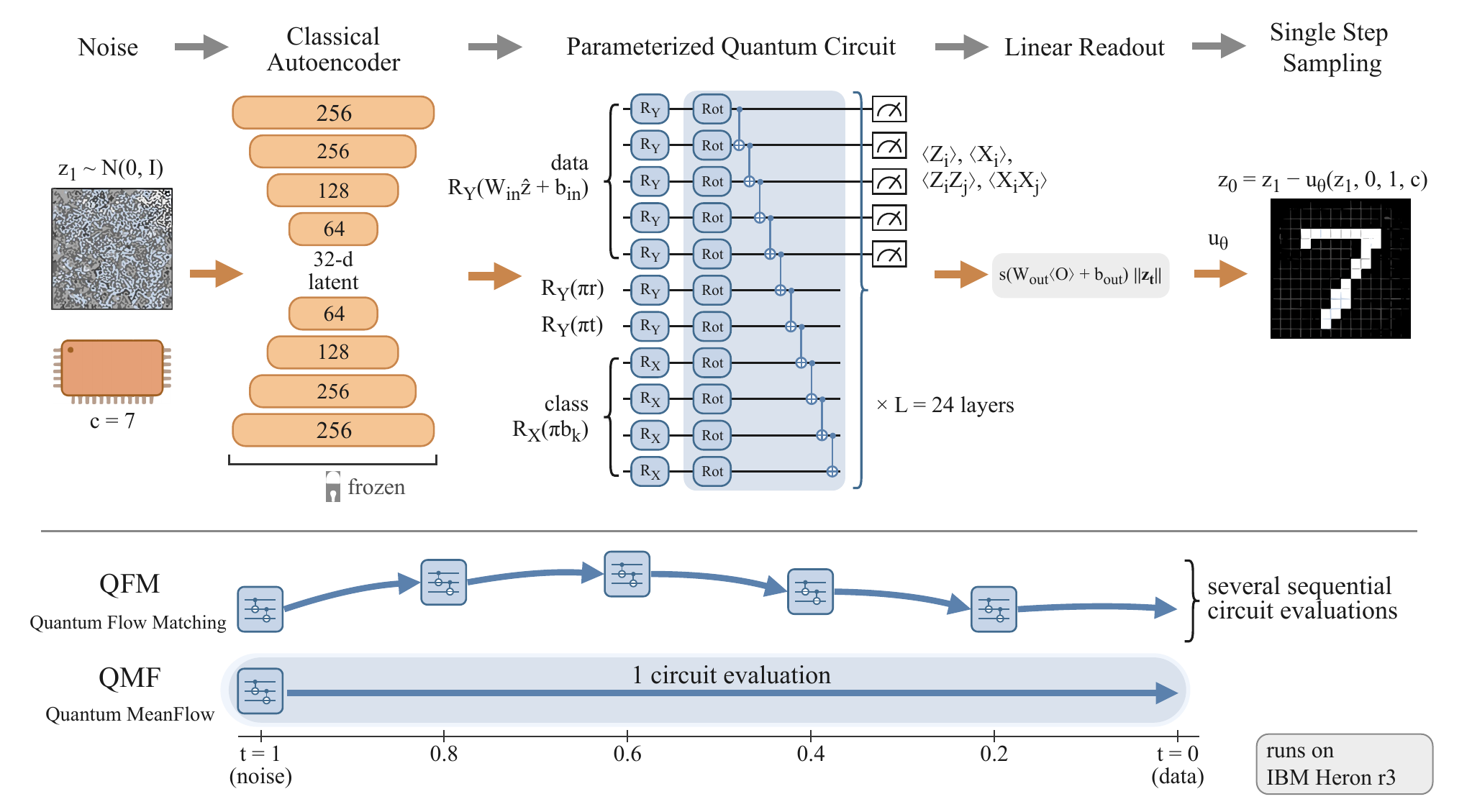}
	\caption{Overview of the quantum flow matching (QFM) and quantum MeanFlow (QMF) models. We train both models on the MNIST dataset. A classical autoencoder provides the data latent vectors, which are the inputs for a parameterized quantum circuit. This circuit learns the velocity fields between the noise and data distribution. QFM requires many sequential evaluations of the trained circuit for sampling while QMF is designed to sample with a single circuit evaluation per measurement basis.}
	\label{fig_overview}
\end{figure}

In this section, we present the quantum implementations of flow matching and MeanFlow, termed as quantum flow matching (QFM) and quantum MeanFlow (QMF). 
For benchmarking, we use the MNIST dataset. The goal was to generate $16 \times 16$ MNIST digits conditioned by class label. 
In the classical case, the velocity field is learned by a neural network. Here, we used a parameterized quantum circuit (PQC) to describe the velocity fields. 
We use a classical autoencoder to assist with the input to the quantum circuit, so our models work not directly on the pixel space but on the latent space. 
We use the same architecture for both QMF and QFM, with QMF having an extra qubit to define the time interval, which means that most of the reported metrics can be attributed to the QMF formulation rather than the differences between architectures and hyperparameters. An overview of the method is depicted in Fig.~\ref{fig_overview}, with the details in the following sections.

%Our goal is to generate $16\times16$ MNIST digits conditioned on a class label by learning a generative velocity field with a parameterised quantum circuit. The field is not learned in pixel space but in the $32$-dimensional latent space of a frozen classical autoencoder, and the quantum circuit is a hardware-efficient ansatz designed to be executable on current devices. We train two models on this common scaffold and compare them throughout: a conditional flow-matching model (QFM), whose sampler requires several sequential circuit evaluations, and the MeanFlow model (QMF), whose sampler requires a single evaluation. The two share an identical training protocol, dataset, and classical scaffold, and QMF differs from QFM only by one extra qubit and its loss and sampler, so that every reported comparison is directly attributable to the MeanFlow formulation rather than to incidental differences in architecture or tuning.

\subsection{Data processing and classical methods}
\label{ss_data}

We worked on the standard MNIST handwritten digits dataset, with ten classes for digits $0-9$ and $60\,000$ training and $10\,000$ test images. 
The original dataset has each image of $28\times28$ pixels. This size is too large for encoding into a quantum computer, especially if we want to avoid amplitude encoding (see section~\ref{ss_ansatz}). We downsampled the original images to $16\times16$ by bicubic anti-aliased interpolation, obtaining grayscale images with pixel values in $[0,1]$. \\

%We use the standard MNIST handwritten-digit dataset ($60\,000$ training and $10\,000$ test images, ten classes). Each $28\times28$ image is downsampled to $16\times16$ by bicubic anti-aliased interpolation and then converted to a half-tone (binary) rendering using void-and-cluster blue-noise dithering \cite{ulichney1993}, which preserves the perceived grey level of the downsampled digit in a $256$-pixel binary pattern; a fixed blue-noise threshold mask is shared across the dataset. The resulting grey images take values in $[0,1]$ and are rescaled to $[-1,1]$ before entering the autoencoder.

We used a classical autoencoder to obtain the latent space representation of the images, which is then encoded into the quantum circuit. 
The autoencoder is a multi layer perceptron (MLP), which takes the flattened 256-dimensional image and converts it to a 32-dimensional latent vector. It is comprised of 4 layers: $256\!\to\!256\!\to\!128\!\to\!64\!\to\!32$ encoder with a mirrored decoder. It uses SiLU activation functions and a tanh output activation, with $\approx\!2.2\times10^{5}$ total parameters. It was trained for 60 epochs with the AdamW optimizer (using a learning rate of $2\times10^{-3}$, weight decay
$10^{-5}$ and cosine-annealing). 
The encoder latent vectors were standardized to zero mean and unit variance. Latent vectors sampled by the PQC were destandardized before entering the decoder.
We also trained a separate MLP classifier which was used for measuring the class accuracy of generated samples and for best-of-$N$ sampling. 
It has 3 layers: $256\!\to\!256\!\to\!128\!\to\!10$, with SiLU activation and achieved $98.2\%$ test accuracy.
The classical machinery is shared across the both QFM and QMF models and is fixed during the quantum training and sampling. 

%A single frozen classical scaffold is shared by every model. An MLP autoencoder compresses a flattened $256$-dimensional image to a $32$-dimensional latent ($256\!\to\!256\!\to\!128\!\to\!64\!\to\!32$ encoder, a mirrored decoder, SiLU activations, a $\tanh$ output, $\approx\!2.2\times10^{5}$ parameters). It is trained for $60$ epochs with AdamW (learning rate $2\times10^{-3}$, weight decay $10^{-5}$, cosine-annealed) to minimise pixel-space MSE, reaching a test reconstruction MSE of $9.3\times10^{-3}$ on the $[-1,1]$ scale. The generative model operates entirely on the encoder latents, which are standardised to zero mean and unit variance per dimension (statistics computed once over the training set) so that the flow-matching loss has a fixed scale; sampled latents are de-standardised before decoding. A separate MLP classifier ($256\!\to\!256\!\to\!128\!\to\!10$, SiLU, $98.2\%$ test accuracy) is trained on the same half-tone images and is used downstream of the generator only, for best-of-$N$ candidate selection and for measuring class accuracy of generated samples. Neither the autoencoder nor the classifier is updated during generative training.

\subsection{Quantum ansatz}
\label{ss_ansatz}

To learn the velocity field, we use a parameterized quantum circuit (PQC), also known as a variational quantum circuit or variational ansatz \cite{benedetti2019pqc,cerezo2021vqa}. 
A PQC is composed of parameterized single-qubit rotation gates interleaved with entangling gates and imposes a differentiable map on a quantum state. 
The output is read out as a set of expectation values of chosen observables. 
The gates can be tuned to minimize some function, similar to how artificial neural networks are trained to minimize the loss function. 
%Parameterised quantum circuits (PQCs), also called variational quantum circuits or ansätze, are quantum circuits whose gates depend on continuous, trainable parameters \cite{benedetti2019pqc,cerezo2021vqa}. Acting on a register of qubits initialised in a fixed reference state, a PQC applies a sequence of parameterised single-qubit rotations interleaved with entangling gates, and the circuit output is read out as a set of expectation values of chosen observables. Each expectation value is a single real number, so the circuit realises a smooth, differentiable map from its inputs to a fixed-length real vector, controlled by the trainable gate angles. This makes PQCs a natural quantum analogue of a neural network layer, and they have become the central building block of quantum machine learning: classical data is written into the circuit through an encoding stage, the variational gates are optimised on a task loss, and the measured outputs are consumed by the surrounding model. PQC-based models have been applied across supervised classification and regression, kernel methods, and generative modelling \cite{mitarai2018qcl,havlicek2019supervised}, and are typically trained in a hybrid quantum–classical loop in which a classical optimiser updates the circuit parameters using gradients evaluated on the quantum device. Because they combine trainable expressivity with shallow, hardware-realisable structure, PQCs are especially suited to the noisy intermediate-scale quantum (NISQ) regime, and we adopt one here to parameterise the velocity field of our flow-matching model.
Here, the PQC is sandwiched between the classical scaffolding described above (see Fig.~\ref{fig_overview}). 
We designed a hardware-efficient ansatz  \cite{kandala2017hea}, with repeating trainable single-qubit rotations and a sparse entanglement, for which sampling can be executed on the current NISQ devices. 
The QMF circuit, depicted in Fig.~\ref{fig_overview}, learns the average velocity $u_\theta(z_t,r,t,c)$. It uses $11$ qubits, with first 5 qubits encoding the data, one qubit for interval initial time $r$, one qubit for interval final time $t$ and 4 qubits for the class labels. 
The QFM circuit is otherwise identical except that it does not contain the qubit corresponding to time $r$ and learns the instantaneous velocity $v_\theta(z_t,t,c)$.\\

We chose different encoding schemes for the data and class labels. 
The 32-dimensional latent vector obtained from the classical encoder was used as the input for the PQC. Since this latent is a continuous-valued vector, we chose angle encoding to encode it into the qubits. The latent vector was first normalized as $\hat z=z_t/\lVert z_t\rVert$, and its magnitude $\lVert z_t\rVert$ is retained for the output. We trained a linear map $W_{\mathrm{in}}\hat z+b_{\mathrm{in}}$ ($W_{\mathrm{in}}\in\mathbb{R}^{5\times32}$) producing five angles which were encoded as $R_Y$ rotations into the 5 data qubits. 
The class labels, being discrete variables, use basis encoding. 
The label $c\in\{0,\dots,9\}$ is expressed in the binary form as $c=\sum_{k=0}^{3} b_k\,2^{k}$ with bits $b_k\in\{0,1\}$. Each class qubit is prepared with a rotation $R_X(\pi b_k)$, so that the qubit $k$ stays $\ket{0}$ when $b_k=0$ and flipped to $\ket{1}$ when $b_k=1$. 
The four class qubits, therefore, can be represented by the computational basis state $\ket{b_3 b_2 b_1 b_0}$.
This maps each class to a distinct, mutually orthogonal state, giving a clean and maximally distinguishable conditioning input.
Finally, the two time points are encoded as rotations $R_Y(\pi t)$ and $R_Y(\pi r)$. For the QFM model, the wire corresponding to time $r$ is dropped. In all cases, we avoided using amplitude encoding as it uses several entangling gates which causes the state-preparation depth to grow exponentially with the number of qubits, making a NISQ implementation difficult. Furthermore, empirically, we found, in simulations, that amplitude encoding resulted in poorer results than the current encoding scheme.\\

%\paragraph{Input encoding.} A latent vector $z_t\in\mathbb{R}^{32}$ is first normalised to unit norm, $\hat z=z_t/\lVert z_t\rVert$, and its magnitude $\lVert z_t\rVert$ is retained for the output. The normalised latent is written onto the five data qubits by \emph{angle encoding}: a trainable linear map $W_{\mathrm{in}}\hat z+b_{\mathrm{in}}$ ($W_{\mathrm{in}}\in\mathbb{R}^{5\times32}$) produces five angles applied as $R_Y$ rotations. The four class bits $b_k$ are prepared with $R_X(\pi b_k)$, and the two times are encoded as $R_Y(\pi t)$ and $R_Y(\pi r)$ on their respective wires. Angle encoding costs a single layer of single-qubit gates and no entangling gates, in contrast to amplitude encoding whose state-preparation depth grows with the input size.

The trainable block of the circuit is composed of $L=24$ layers, designed in a linear-chain topology so that it maps directly onto the IBM Heron R3 devices without routing overhead. Each layer consists of a single-qubit rotation gate on every qubit followed by a nearest neighbor CNOT chain, resulting into 10 CNOTs per layer for the QMF circuit (240 two-qubit gates) and 9 CNOTs per layer for the QFM circuit (216 two-qubit gates). The output is obtained as Pauli expectation values on the five data qubits in the $Z$ and $X$ basis. We used the single qubit expectation values $\langle Z_i\rangle$ and
$\langle X_i\rangle$ as well as the pairwise expectation values $\langle Z_iZ_j\rangle$ and
$\langle X_iX_j\rangle$, resulting in a 30-dimensional vector. This is mapped to a 32-dimensional latent space using a trainable classical linear readout (weights and biases as $W_\text{out}$ and $b_\text{out}$, respectively), scaled by a single trainable scalar $s$. 
After multiplying with the magnitude $\lVert z_t\rVert$, the full forward map for the QMF model is obtained as 
\begin{equation}
u_\theta(z_t,r,t,c)=\lVert z_t\rVert s\bigl(W_{\mathrm{out}}\langle\mathbf O\rangle(\hat z,r,t,c;\theta) + b_{\mathrm{out}}\bigr),
\end{equation}
where $\langle\mathbf O\rangle$ denotes the measured expectation values. Note that $s$ serves to scale the output so that the PQC supplies the direction of the velocity in the latent space, while its magnitude is carried by $\lVert z_t\rVert$ classically. This results in more stable training. The trainable parameters in the quantum circuits are $L\times 3\times n_{\mathrm{qubits}}$, i.e, $792$ for QMF and $720$ for QFM.

\begin{figure}
        \centering
        \includegraphics[width=1\linewidth]{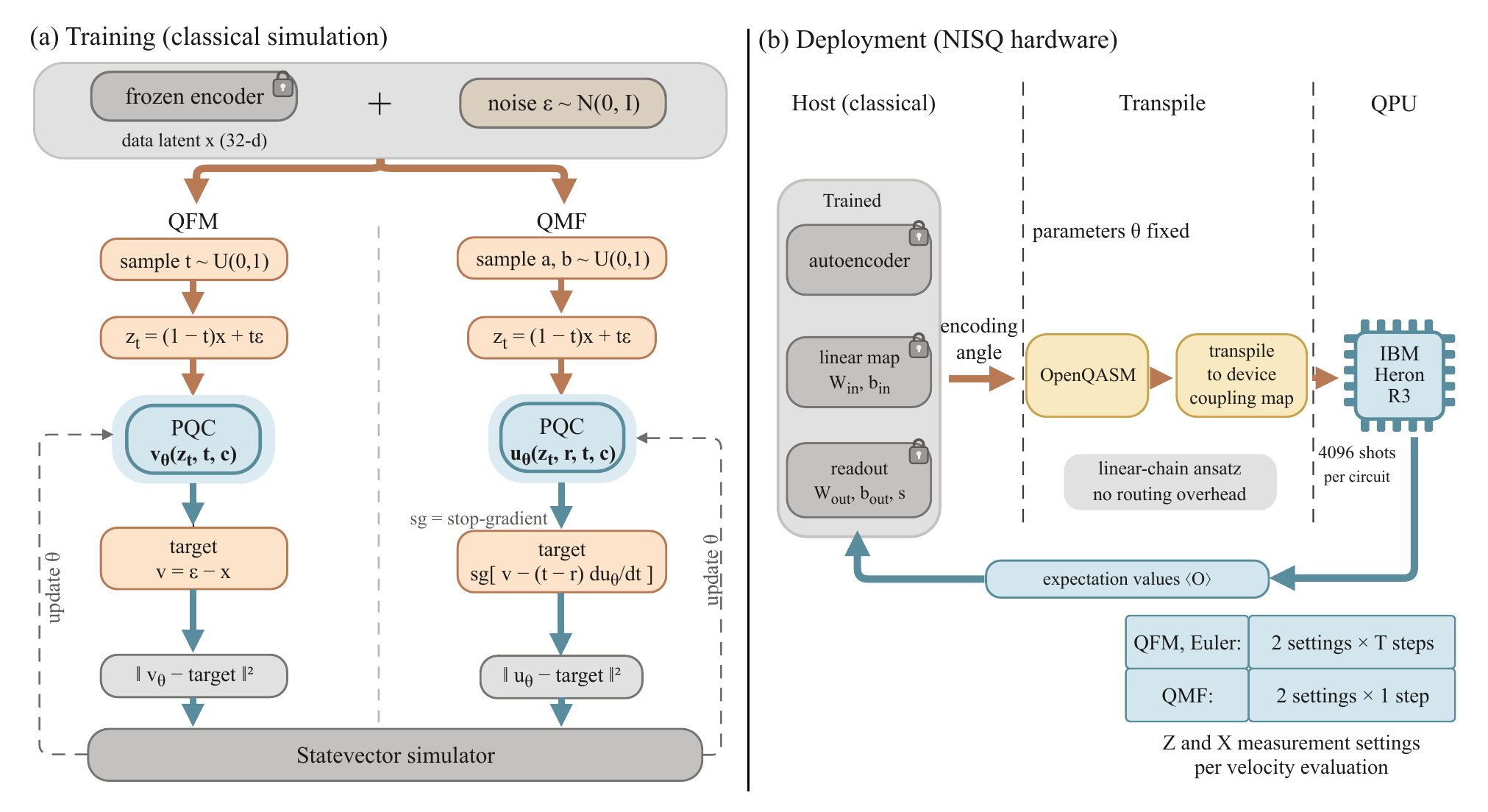}
	\caption{Training and sampling pipeline. (a) Training procedure for the QFM and QMF models. Both models share the same classical backbone, with an autoencoder providing the data latent vector. The quantum circuit in QFM learns the instantaneous velocity while in QMF, it learns the average velocity. (b) Once the models are trained, sampling can be performed on QPU.}
	\label{fig_training}
\end{figure}

\subsection{Training and sampling}
\label{ss_training}

Both models were trained classically using PennyLane's \texttt{default.qubit} simulator \cite{bergholm_pennylane_2022}. We used a statevector simulator as training on real hardware for these models is still not feasible due to high error rates. We used the AdamW optimizer with a learning rate of $3\times10^{-3}$, weight decay of $10^{-5}$ and cosine annealing. We also used gradient-norm clipping at $1.0$ and trained for 30 epochs. For the QFM model, we minimized the conditional FM loss given in Eq.~\eqref{eq:fm}. That is, for a data latent vector $x$ and noise $\varepsilon\sim\mathcal N(0,I)$, we sampled $t\sim\mathcal U(0,1)$ and formed $z_t=(1-t)x+t\varepsilon$. The network output is then regressed against the target velocity $v=\varepsilon-x$.\\

%Both models are trained classically by simulating the circuit exactly on a statevector backend (PennyLane \texttt{default.qubit}), so gradients flow through the quantum layer by automatic differentiation; on hardware the same gradients would be obtained through the parameter-shift rule. All training uses AdamW (learning rate $3\times10^{-3}$, weight decay $10^{-5}$, cosine-annealed over the run), gradient-norm clipping at $1.0$, $30$ epochs over the $60\,000$ latents, a fixed random seed, and an exponential moving average (EMA, decay $0.999$) of the weights that is used for all sampling and evaluation.

%For QFM we minimise the conditional flow-matching loss of Eq.~\eqref{eq:fm}: for a data latent $x$ and noise $\varepsilon\sim\mathcal N(0,I)$ we draw $t\sim\mathcal U(0,1)$, form $z_t=(1-t)x+t\varepsilon$, and regress the network output onto the constant target $v=\varepsilon-x$. Batches contain $128$ samples.

For the QMF model, we minimized the MF identity loss given in Eq.~\eqref{eq:mfloss}. Here, for each data latent vector $x$ and noise sample $\varepsilon\sim\mathcal N(0,I)$, we sampled two independent quantities $a,b\sim\mathcal U(0,1)$ and set $r=\min(a,b)$ and $t=\max(a,b)$. Then, forming $z_t=(1-t)x+t\varepsilon$, we regress the network output against the target $\mathrm{sg}[\,v-(t-r)\,\mathrm{d}u_\theta/\mathrm{d}t\,]$. To calculate the target, the total time derivative of $u_\theta$ along the whole trajectory is required. This is obtained using a single Jacobian-vector product of $u_\theta$ at $(z_t,r,t)$ with the tangent $(v,0,1)$. Compared to the original formulation in \cite{geng2025MeanFlow} we simplified our training in two ways. First, we drew $r$ and $t$ as the minimum and maximum of the two uniforms rather than having a large fraction at $r=t$, and second, we used an unweighted square error instead of their adaptive weighting. These choices are based on their superior empirical performance in the QMF model.\\

%For QMF we minimise the MeanFlow-identity loss of Eq.~\eqref{eq:mfloss}. For each data latent we draw a noise sample, draw two independent uniforms $a,b\sim\mathcal U(0,1)$ and set $r=\min(a,b)$, $t=\max(a,b)$, and form the interpolant $z_t=(1-t)x+t\varepsilon$ anchored at the upper time $t$ (following the convention $t=0$ data, $t=1$ noise of Sec.~\ref{ss_flowmatching}). The target $\mathrm{sg}[\,v-(t-r)\,\mathrm{d}u_\theta/\mathrm{d}t\,]$ requires the total time derivative of $u_\theta$ along the trajectory, which we obtain as a single forward-mode Jacobian-vector product of $u_\theta$ at $(z_t,r,t)$ with tangent $(v,0,1)$; the value and its directional derivative are returned in one pass. The JVP is taken with \texttt{create\_graph} enabled so that back-propagation still reaches the circuit parameters through the left-hand $u_\theta$, while the right-hand target is detached (the stop-gradient). Because each JVP costs roughly three times a plain forward/backward step, we halve the batch to $64$ relative to QFM. We depart from \cite{geng2025MeanFlow} in two simplifications, both motivated empirically in Sec.~\ref{s_results}: we draw the two times as $\min/\max$ of two uniforms rather than anchoring a large fraction at $r=t$, and we use an unweighted squared error in place of their adaptive weighting. On CPU, one QMF epoch takes $\approx\!20$ minutes and the full run $\approx\!10$ hours (roughly $3\times$ the matched QFM run), consistent with the JVP overhead.

Once the models are trained, new samples can be generated by producing new latent vectors that are decoded using the frozen autoencoder. 
For the QFM model, generation is carried out by integrating the instantaneous velocity field from noise $z\sim\mathcal N(0,I)$ to data via an integrator of choice. In section~\ref{s_results} we compare the Euler and Heun integrators for several time steps. For the QMF model, single step sampling is achieved by drawing $z_1\sim\mathcal N(0,I)$ and using the average velocity to obtain $z_0=z_1-u_\theta(z_1,r{=}0,t{=}1,c)$. \\

%Generation always operates on latents and decodes the result with the frozen autoencoder. For QFM, sampling integrates the learned instantaneous velocity from noise to data: starting from $z\sim\mathcal N(0,I)$ we take a fixed number of ODE steps (a $60$-step Heun integrator in simulation, i.e.\ two network evaluations per step and $120$ evaluations per image; a shorter Euler schedule is used on hardware, see Sec.~\ref{ss_hardware}). For QMF, sampling is one step: we draw $z_1\sim\mathcal N(0,I)$ and set $z_0=z_1-u_\theta(z_1,r{=}0,t{=}1,c)$ — a single circuit evaluation per image. A multi-step MeanFlow sampler is also available for diagnostics — descending $t$ from $1$ to $0$ and subtracting $(t-r)\,u_\theta$ at each step — where every step is exact under the identity rather than an Euler approximation; for a well-fitted model the single-step and multi-step outputs coincide, and the residual gap between them is our measure of how well the average velocity has been learned.

%Because MNIST samples from a per-draw velocity field are individually imperfect, we report both a default (single-candidate) accuracy and a \emph{best-of-$N$} accuracy, in which $N$ independent trajectories are generated per requested class and the frozen classifier keeps the candidate with the highest log-probability of the target class. Best-of-$N$ is inexpensive for QMF because each candidate is a single evaluation; we use $N=8$ for QFM and up to $N=32$ for QMF. The selection rule is identical across models so best-of-$N$ numbers are directly comparable.

One of the main objectives of this study is to create a quantum generative model that can be implemented on the current NISQ devices. While training the two models on hardware is still not feasible, inference using the trained models is possible. 
The classical machinery is executed on the host device and each velocity evaluation takes place on the QPU. 
For each sampling step, the trained circuit is instantiated with the encoding angles, exported to OpenQASM and transpiled to the specific device's coupling map. 
As the measurements are in two different bases, for the QFM model, two jobs (four jobs) are required per Euler step (Heun step), which makes total circuit evaluations scale with number of steps. For the QMF model, two jobs is required to generate one image. 
We found that at this circuit depth, error mitigation techniques like zero-noise extrapolation do not provide any gain.

\section{Results}
\label{s_results}

\begin{figure}
        \centering
        \includegraphics[width=1\linewidth]{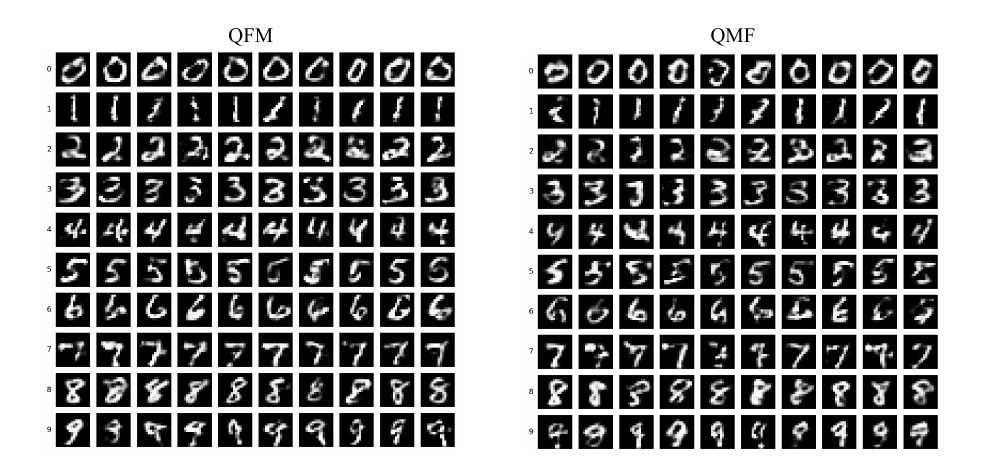}
	\caption{Samples generated by the QFM model (left) and the QMF model (right) with statevector simulation. For QFM sampling, we used 60 time steps and Heun integrator, while the QMF sampling is done with a single step.}
	\label{fig_results1}
\end{figure}

In this section, we present the results for both the QFM and QMF models. 
After training, the samples can be generated either on a simulator or on a quantum processing unit (QPU). 
In statevector simulations, the QFM model (with 60 time steps and Heun integrator) generates the correct digit for a given condition with an accuracy of 0.85, while the QMF model (single-step) has an accuracy of 0.63. 
Since single draws vary in quality because of stochastic noise seed and approximate velocity field, a cheap classifier-based rejection sampling, known as best-of-$N$ (BoN) sampling, can be used to recover accuracy and generate high quality samples without modifying the model. Under the BoN sampling, with $N=8$, both models saturate to an accuracy of 1.0. The samples generated under BoN sampling for both models are shown in Fig.~\ref{fig_results1}.
The quality of the generated images is characterized using the Fréchet distance (referred to as FID from now) between the real and generated image distributions, evaluated in the feature space of the trained classifier rather than of the usual Inception network. Let $(\mu_r,\Sigma_r)$ and $(\mu_g,\Sigma_g)$ denote the mean and covariance of the classifier features of the real and generated images, respectively. The FID score is defined as 
\begin{equation}
\mathrm{FID}=\lVert\mu_r-\mu_g\rVert^2+\mathrm{Tr}\!\left(\Sigma_r+ \Sigma_g - 2  \left( \Sigma_r \Sigma_g \right)^{1/2}\right).
\end{equation}
We have used 2000 samples for each FID calculation. Using classifier features for real images only gives us an FID floor at 1.89.
The single-sample, before BoN rejection sampling, FID of the images generated by the QFM model is 10.62 while that of the QMF is 28.26. For reference, the respective FIDs before training was 125.6 and 150.3. This shows that training of the quantum circuit substantially improves the generated image quality. The lower single-sample FID for the QMF model is the trade off made to achieve single-step sampling.\\

\begin{figure}
        \centering
        \includegraphics[width=1\linewidth]{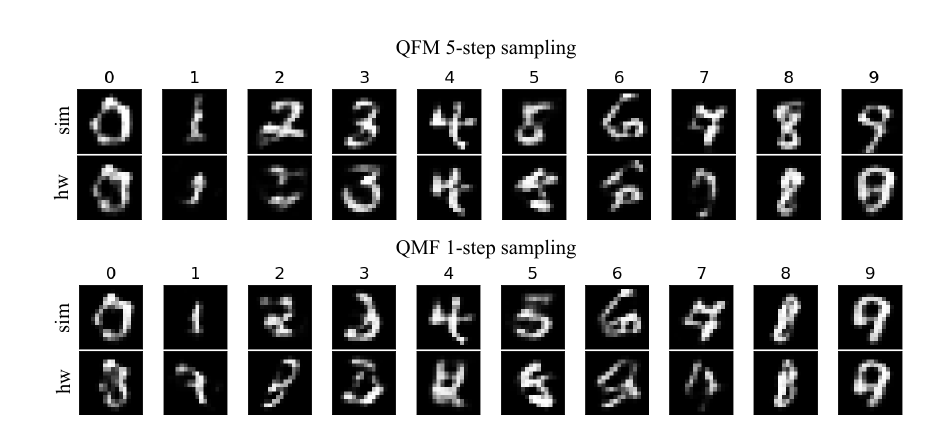}
	\caption{Samples generated on QPU compared with those obtained from the simulator. (a) QFM samples with 5 time steps and Euler integrator. (b) QMF single-step sampling. In both (a) and (b), top row shows samples generated on the simulator (sim) and the bottom row shows samples generated on the quantum hardware (hw).}
	\label{fig_results2}
\end{figure}

We next demonstrate inference on a QPU. We evaluated the trained models on two IBM machines: Heron r2  \texttt{ibm\_pittsburgh} and Heron r3 \texttt{ibm\_boston} machine with 4096 shots per measurement basis. 
For the QFM model, each time step requires one circuit evaluation per measurement basis as the output of the current step is used as the input to the next step. This makes the QPU cost very high. To avoid this, we used only 5 time steps and changed the integrator from Heun to Euler to halve the circuit evaluations per time step. 
The Pearson correlation between the hardware output and ideal values for \texttt{ibm\_boston} (\texttt{ibm\_pittsburgh}) is 0.57 (0.26) for QFM and 0.23 (0.10) for QMF. This is because of higher gate errors in the Heron r2 machine. For the rest of this section, we report only Heron r3 results. 
Fig.~\ref{fig_results2} shows the samples generated on the \texttt{ibm\_boston} QPU and compares them with those generated on the simulator for both the models. 
As expected, due to gate and readout noise, the quality of hardware generated images is poorer than that of simulator. 
The single-sample accuracy also falls to 0.30 and 0.20 for QFM and QMF, respectively.
BoN sampling strategy, with $N=8$, raises these numbers to 0.80 for QFM and 0.60 for QMF on the QPU, and brings them close to the single-sample accuracy on the simulator. This shows the advantage of the BoN strategy on hardware: without changing the models and circuits we can obtain accurate samples from several independent noisy trajectories. It masks the hardware error with the tradeoff of $N$ independent runs per image. \\

\begin{figure}
        \centering
        \includegraphics[width=1\linewidth]{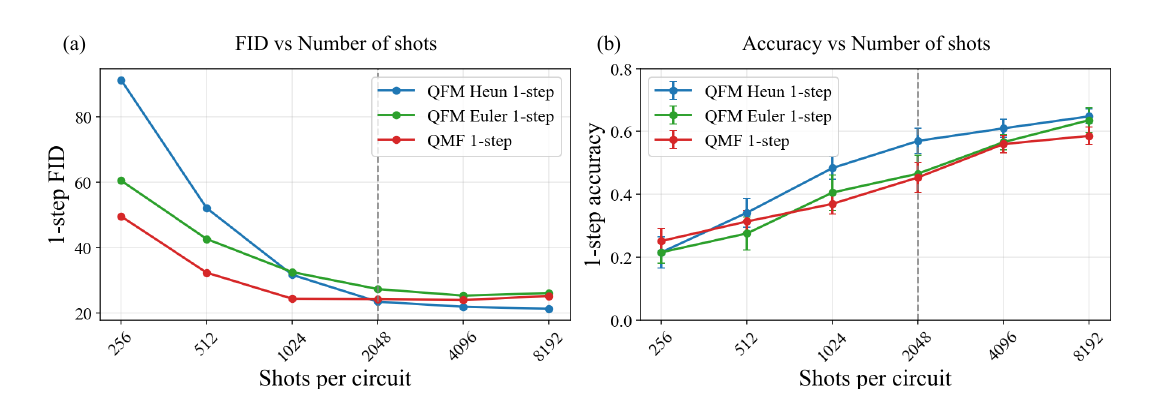}
	\caption{Single-step sampling performance as a function of the number of shots. (a) FID and (b) accuracy of both the QFM and QMF models. The dashed vertical line indicates the number of shots when single-step QFM using Heun's method (two circuit evaluation per step per measurement basis) begins to generate higher quality images than single-step QMF sampling.}
	\label{fig_results3}
\end{figure}

Classically, the main advantage of using a QMF model is the single-step sampling, which saves on the cost of integrating an ODE over several time steps. 
Although the QMF model generates lower quality images compared to the QFM model  (28.26 vs 10.62 FID), under a finite shot budget we have a different picture.
Here, we compare single step FIDs and accuracies for the two models with number of shots, shown in Fig~\ref{fig_results3}. For the QFM model, we compare both integrators. These experiments were performed in simulations. 
In all cases, the quality and accuracy gets better with number of shots, as expected because of decreasing sampling noise. 
For low number of shots, the QMF model has better single-step FID than the QFM model for both integrators. At 2048 shots, Heun integrator starts performing better. However, Heun integrator is a two-step method for each time step and raises the cost of inference two times.
The QMF model always outperformed the QFM with Euler integrator, which has one circuit evaluation per time step.
For accuracy (5 seeds $\times$ 100 samples), QFM Euler and QMF agree within error bars, while QFM Heun generally performs better. \\

\begin{figure}
        \centering
        \includegraphics[width=1\linewidth]{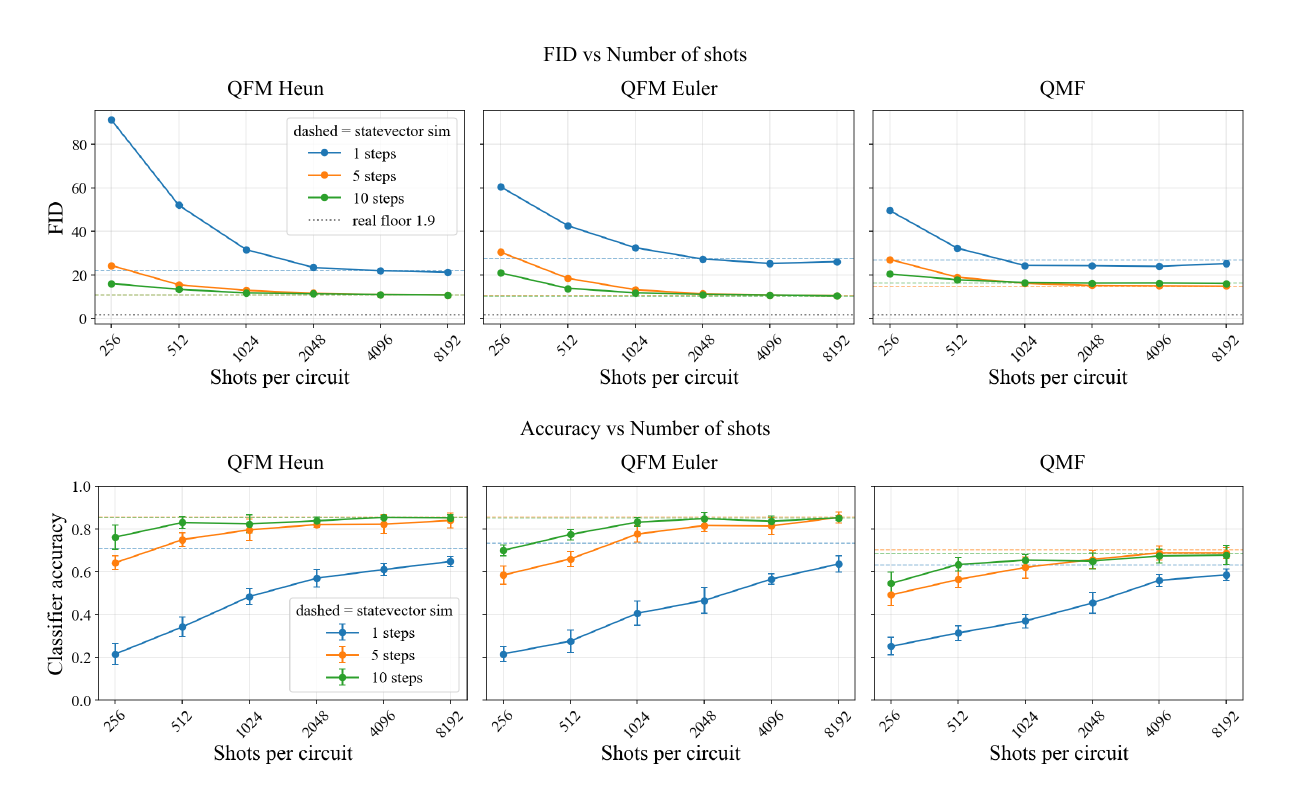}
	\caption{Multi-step sampling performance metrics with number of shots. Top row shows the FID and bottom row shows the accuracy. The colored horizontal lines show the FID and accuracy values obtained on the statevector simulator and the dashed grey line shows the minimum value possible for FID.}
	\label{fig_results4}
\end{figure}

If we relax the condition of single-step generation, we can obtain higher quality images. Here, we study the tradeoff between the image quality and number of sampling steps. Fig.~\ref{fig_results4} shows FID and accuracies for multi-step sampling of the both models with number of shots. 
Increasing the number of time steps results in an increase in performance not only for the QFM model but also for the model QMF. This is possible if the average velocity is not learned properly by the QMF model, see Sec.~\ref{s_discussion}. 
However, the most interesting behavior across all plots is that increasing the number of time steps increases the quality and accuracy even at very small number of shots. For example, for the QFM model with Heun integrator, FID at 256 shots falls from 91.2 at 1 step to 16.1 at 10 steps. 
This means that an increase in integrator time steps buys robustness to sampling noise. 
For all models, moderately high shots give similar performance as in the statevector simulations.
Furthermore, for the QFM model at moderately high shots, 5 and 10 time steps have almost the same performance (FID $\approx$ 10.5), which is never reached by QFM single step sampling. Finally, the QMF model, even at multiple steps, does not reach the same accuracy or FID as the QFM model. 

%Parameter efficiency vs classical? make the plot tomorrow and decide.\\ No because MeanFlow classical performs better.
% (maybe in v2) Finally, we study generation diagnostics. \\
%2. Parameter efficiency versus classical. (ready)\\

\section{Discussion}
\label{s_discussion}

Here, we demonstrate the first application of the MeanFlow formulation, which enables high-quality single-step sampling in classical flow matching, for quantum flow matching (QFM), which we termed quantum MeanFlow (QMF). 
We achieved this by using a parameterized quantum circuit in place of a neural network used in the classical scheme, which learns the velocity field between the two probability distributions. We tested our method on the MNIST database, comparing the performance of the QMF model against the baseline QFM model. 
QMF is a viable single-step generator and delivers better performance than QFM for single-step sampling. This performance is especially pronounced at lower number of shots, where the QFM performance degrades sharply. 
This can be advantageous for tasks that prioritize fast sampling over per-sample fidelity, such as rapid previewing, large-scale candidate generation for downstream filtering, or latency-constrained settings where each additional circuit submission incurs queue and readout overhead. 
We further observed that at lower number of shots we can obtain quality samples by sampling QFM with multiple time steps: 5-step sampling with the Heun integrator at 256 shots (total shots $5\times2\times 256=2560$) gives similar quality and accuracy as single-step Heun sampling at 8192 shots (total shots $2\times 8192=16384$). 
This property should be seen as robustness to shot noise and does not necessarily mean a reduction in hardware cost. This is because each additional step adds a sequential submission to the hardware which has some fixed overhead and latency, typically outweighing the cost of extra shots on the current NISQ devices.
Thus, using a larger shot budget for a single circuit would be preferable over distributing a smaller shot budget across many sequential circuits. This is precisely the regime in which single-step sampling can be advantageous.
\\

The QMF model learns the average velocity over a time interval, while the QFM model learns the instantaneous velocity at a single time step. 
The former is substantially harder to learn because of several reasons. First, the average velocity $u_\theta(z_t,r,t)$ is higher dimensional than the instantaneous velocity $v_\theta(z_t,t)$. 
Second, the loss function of QMF, $v-(t-r)\mathrm{d}u_\theta/\mathrm{d}t$, is self-referential, i.e., it depends on the model's own derivative, which is more complex than the loss function of QFM, $\varepsilon - x$, which is a fixed vector. Finally, QMF generates the whole trajectory at once, which is more demanding than QFM where many small steps are integrated by a classical ODE solver.
The multi-step sampling plot in Fig.~\ref{fig_results4} shows that the performance of the QMF model increases with number of time steps. This implies that the model did not learn the average velocity exactly, because otherwise the single-step and multi-step sampling would have the same performance.
We attribute this performance gap to the limited capacity of the quantum circuit. 
The QMF model may benefit from a more expressive quantum circuit, such as circuits with greater depth, richer entanglement, additional data or time re-uploads. 
However, such quantum circuits may not be executable on the current hardware reliably. Thus, a more capable QMF model that can still run on the current NISQ devices is an important direction for future work. 
Another interesting observation was that the QFM performance saturates just at 10 integrator time steps. This is potentially due to the MNIST dataset being a comparatively easy target for which few integration steps suffice. 
Testing both models for more complex and larger datasets will also be an important direction for the future. Lastly, as we found with our experiments with the IBM machines, hardware with better fidelity and deeper circuits would enable generation of higher quality data from these quantum generative models.

%measuring all qubits or different sets of observables does not produce huge changes in the results.
%1. MeanFlow low accuracy, performance capped and difficulty in learning average velocity.\\
%2. QFM working well with small number of steps likely due to easier dataset.\\
%An advantage of our method is that the sampling can be on a real device. Shots are expensive so we may use QMF. \\

\section{Conclusion}
\label{s_conclusion}

In this study, we have introduced Quantum MeanFlow (QMF), the first application of the classical MeanFlow formulation to quantum flow matching (QFM). 
In QFM, a parameterized quantum circuit (PQC) is trained to learn the instantaneous velocity field which is integrated over several time steps by an ODE solver for sampling. 
QMF enables single-step sampling as the PQC learns the average velocity over a time interval instead.
We benchmarked both models on the MNIST handwritten digits database.
Both of our models were trained in simulations, though inference was performed on real quantum hardware. 
QFM requires several steps for sampling which results in increased hardware costs due to sequential input and output. 
QMF offers a viable alternative for single-step sampling as it consistently performs better than single-step QFM sampling. 
When executed on a QPU, QMF requires only a single circuit evaluation (per measurement basis) for an image, and the loss in accuracy caused by device noise can be mitigated using best-of-$N$ rejection sampling.
The main limitation of the current QMF model is that the average velocity is an intrinsically harder target to learn than the instantaneous velocity and the single-step sample quality is capped by the capacity of quantum circuit executable on NISQ devices.
Development of more expressive circuits that can fit on the NISQ devices and extending these models to more complex data are directions for future work.
Taken together, our results show that QMF can substantially reduce the quantum computational cost of generative sampling while maintaining competitive performance on current quantum hardware. 
More broadly, this work demonstrates that reformulating the generative process itself, rather than relying solely on improvements in hardware or error mitigation, can make quantum generative models more viable under the constraints of near-term devices. This provides a path toward increasingly practical and scalable quantum generative modeling as quantum hardware continues to improve.

\section{Acknowledgements}
\label{s_ack}

Human Biology-Microbiome-Quantum Research Center (Bio2Q) is supported by World Premier International Research Center Initiative (WPI), MEXT, Japan.
This work was also supported by the Center of Innovation for Sustainable Quantum AI (JST Grant Number JPMJPF2221). The authors are grateful to Scott Behie for valuable discussions and comments on the manuscript.

\bibliography{references}

\end{document}